# Experimental determination of the frequency and field dependence of Specific Loss Power in Magnetic Fluid Hyperthermia

M. Cobianchi[1], A. Guerrini[2], M. Avolio[1], C. Innocenti[2], M. Corti[1], P. Arosio[4], F. Orsini[4], C. Sangregorio[3] and A. Lascialfari[4,1]

[1] *Dipartimento di Fisica and INSTM, Università degli Studi di Pavia, Via Bassi 6, 27100, Pavia, Italy;*
[2] *Dipartimento di Chimica, Università degli studi di Firenze, Via della Lastruccia 3, Sesto F.no (FI), Italy;*
[3] *CNR-ICCOM, Via della Lastruccia 3, Sesto F.no (FI), Italy;*
[4] *Dipartimento di Fisica and INSTM, Università degli Studi di Milano, Via Celoria 16, 20133 Milano, Italy;*

Magnetic nanoparticles are promising systems for biomedical applications and in particular for Magnetic Fluid Hyperthermia, a promising therapy that utilizes the heat released by such systems to damage tumor cells. We present an experimental study of the physical properties that influences the capability of heat release, i.e. the Specific Loss Power, *SLP*, of three biocompatible ferrofluid samples having a magnetic core of maghemite with different core diameter $d$ = 10.2, 14.6 and 19.7 nm. The *SLP* was measured as a function of frequency $f$ and intensity of the applied alternating magnetic field $H$, and it turned out to depend on the core diameter, as expected. The results allowed us to highlight experimentally that the physical mechanism responsible for the heating is size-dependent and to establish, at applied constant frequency, the phenomenological functional relationship $SLP = c \cdot H^x$, with $2 \leq x < 3$ for all samples. The *x*-value depends on sample size and field frequency/ intensity, here chosen in the typical range of operating magnetic hyperthermia devices. For the smallest sample, the effective relaxation time $\tau_{eff} \approx 19.5$ ns obtained from *SLP* data is in agreement with the value estimated from magnetization data, thus confirming the validity of the Linear Response Theory model for this system at properly chosen field intensity and frequency.

I.  INTRODUCTION

Magnetic nanoparticles are promising tools in biomedical applications against cancer, and suitable systems for diagnostics by e.g. Magnetic Resonance Imaging and innovative therapies, like drug delivery and Magnetic Fluid Hyperthermia (MFH) [1-11]. The MFH is a recently developed anti-cancer locally acting technique which aims to reduce the side effects of the traditional techniques as chemo- or radio- therapies [12]. This technique makes use of the capability of magnetic nanoparticles (MNPs) to release heat when exposed to an alternating magnetic field (AMF), as a therapeutic treatment to selectively destroy tumor cells within the human body. In MFH treatments, the AMF application is strictly limited to a safety range of frequency $f$ and intensity $H$ due to medical and technical restrictions, as established by the Brezovich criterion which requires $H \cdot f < 4.85 \cdot 10^8$ Am$^{-1}$s$^{-1}$ [13]. The amount of magnetic field energy converted into heat (and subsequently absorbed by tissues) is given by the energy losses occurring during the MNP magnetization reversal, according to the dynamic regime determined by the MNPs physico-chemical properties (size, shape, kind of magnetic core, etc.) as well as by the external field and temperature [14-18]. Magnetization reversal can be produced by a rigid rotation of the particles (Brown process) or by a coherent re-orientation of the electronic spins inside the particles (Néel process) [19-22]. The characteristic times of such processes are respectively $\tau_B = 3\eta V_H/k_B T$, where $\eta$ is the viscosity of the environmental fluid, $V_H$ the hydrodynamic volume of the particle, $k_B$ the Boltzmann constant and $T$ the temperature of the system, and $\tau_N = \tau_0 exp(KV/k_B T)$, where $K$ is the magnetic anisotropy constant of the material, $V$ the volume of the magnetic core and $\tau_0$ the attempt time typically assumed to be $10^{-9}$ s [3,23,24].

The parameter that describes the MNPs capacity to release heat to the surrounding environment when exposed to an AMF is the Specific Loss Power (*SLP*). In Fig.1 the most common empirical models used to evaluate the *SLP* are reported in their intervals of validity [25,26], established on the basis of the MNPs core diameter (whose value depends on the material) and the magneto-thermal quantity $\xi = \mu_0 M_s V H_{max}/k_B T$, where $M_s$ is the magnetic saturation, and $H_{max}$ the maximum applied field. The dimensionless parameter $\xi$ was introduced by Carrey [27] to indicate the limit of validity of the Linear Response Theory (LRT), which holds when $M$ varies linearly with $H$. Indeed, when $\xi << 1$, it is known that the Langevin function, which gives the relationship between $M$ and $H$, can be approximated to a linear function of $\xi$. In this condition one has $SLP \propto H^2$, as we will detail in the following. When $\xi > 1$, a non-linear contribution to the Langevin function must be considered and the Rayleigh model, which foresees $SLP \propto H^3$, is applied [28]. Even though the Rayleigh model has been introduced for ferromagnetic bulk material (multidomain state of Fig. 1), it has been shown that the law $SLP \propto H^3$ could describe the hysteresis losses also in the ferromagnetic (FM) single-domain systems, in the case of coherent reversal. In the single domain state, indeed, the LRT can be properly applied only in the superparamagnetic (SP) regime, while in the FM or blocked regime the *SLP* can be evaluated by using the Stoner-Wohlfarth model (SWM) if the condition $\mu_0 H_{max} > 2\mu_0 H_c$ is satisfied, $H_c$ being the coercive field of the MNPs. However, since $H_c$ depends on the volume of the MNPs, there is a volume range, formally in the FM regime [29,30], where the LRT model still holds (oblique line in Fig. 1). Therefore, in this region, the *SLP* evaluation is not clearly supported by a single model and thus also the exponent which describes the *SLP* variation with $H$ is not univocally foreseen.

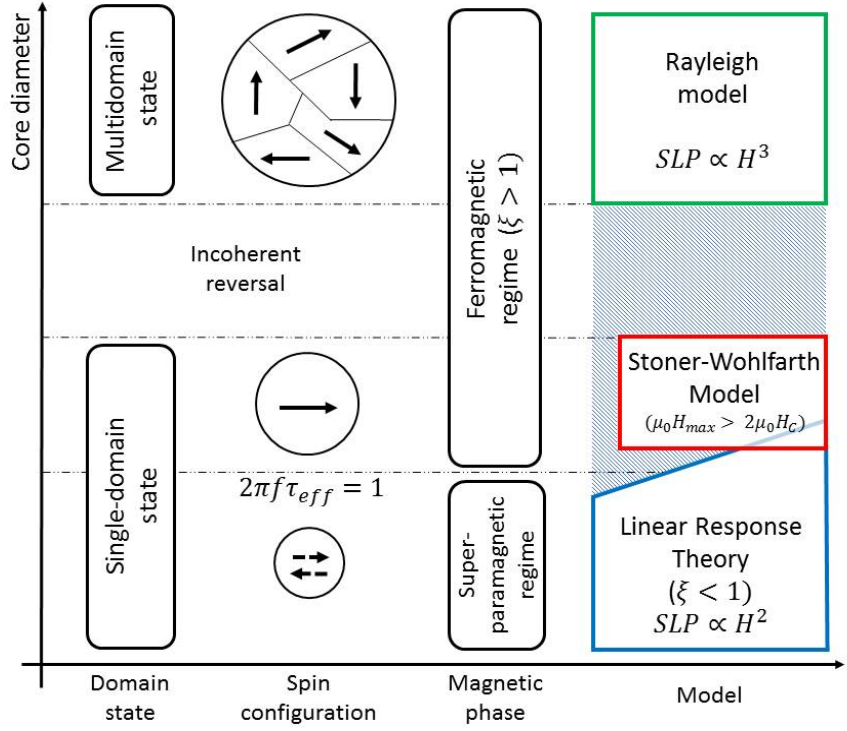

FIG. 1. Schematic representation of the empirical models which describe the *SLP* of MNPs in their intervals of validity, as determined by the core diameter and the magnetic phase (ferromagnetic or superparamagnetic) of MNPs. The quantity $\xi$ depends on extrinsic ($H,T$) and intrinsic ($V, M_s$) parameters (see main text). The transition between superparamagnetic and ferromagnetic (or, alternatively, unblocked/blocked) regimes is here identified with the frequency-dependent condition $2\pi f \tau_{eff} = 1$. The figure was adapted from ref. [27].

Nowadays, most of the literature about MFH assumes the validity of the LRT model often without estimating $\xi$; as a consequence the LRT can be improperly applied. For the outlined reasons, a systematic experimental investigation of the *SLP* behavior in the field and frequency ranges suitable for MFH application is required, to establish the ideal conditions for novel MNPs therapeutic applications. Moreover, a complete set of *SLP(H,f)* experimental data is the base to achieve a deeper understanding of the physical mechanisms underlying the heating processes.

In this work, we present a collection of *SLP* data on three novel maghemite ($\gamma$-Fe$_2$O$_3$) MNP samples with different magnetic core diameters ($d$ =10.2, 14.6 and 19.7 nm, see also Table 1, below) coated with PolyAcrylic Acid (PAA). The data have been collected as a function of the AMF frequency $f$ and amplitude $\mu_0 H$, both chosen in the range of values typically used for in-vitro, pre-clinical and clinical MFH experiments. We suggest also the physical mechanism responsible for the heat release in different $f$ and $\mu_0 H$ intervals, discussing the experimental data within the existing models framework.

## II. EXPERIMENTAL SECTION: MATERIALS AND METHODS

The three samples of $\gamma$-Fe$_2$O$_3$-based MNPs were synthesized by thermal decomposition of metal-organic precursors in high boiling solvents, in the presence of surfactant. Iron(III) acetylacetonate (2 mmol),

Oleylamine (2,5 mmol) and Oleic Acid (2 mmol) were dissolved in Benzylether (40 mL) and stirred under nitrogen flow for 15 min at room temperature; the mixture was heated at 200°C for 30 min and then maintained at 300°C for a variable duration to obtain MNPs of three different sizes. Then, the mixture was cooled down to room temperature, the black MNPs precipitate was magnetically separated, cleaned with ethanol and suspended in toluene. Finally, in order to exchange the Oleic Acid coating and suspend the MNPs in water, 4 mL of each sample, with MNPs concentration 10 mg/mL, was reacted with 40 mg of PolyAcrylic Acid (PAA) in TetraHydroFuran (THF).

The compositional, structural and magnetic properties of the samples were investigated by means of Transmission Electron Microscopy (TEM), Atomic Force Microscopy (AFM) and SQUID Magnetometry, in addition to X-ray diffractometry (data not reported).

The size and morphology of the MNPs cores were determined by TEM, using a CM12 PHILIPS microscope operating at 100 kV. The AFM images were collected using a Bruker Nanoscope Multimode IIId system operating in air, in tapping-mode. Rectangular silicon probes with nominal spring constant around 2.5 N/m (NSG01, NT-MDT) and cantilever length of 120 μm were used. The cantilever resonance frequency was about 130 kHz.

Magnetic measurements were performed by means of a SQUID Quantum Design MPMS magnetometer. Magnetization curves were collected as a function of temperature in the range 2-300 K, by applying a weak magnetic field ($\mu_0 H$ = 5 mT) after cooling the sample in the absence (ZFC, zero-field cooling) and in the presence (FC, field cooling) of the probing field. The hysteresis loops have been collected at $T$ = 2.5 K and $T$ = 300 K in the range -5÷5 T.

## III. RESULTS AND DISCUSSION

In Table I the mean diameters of the three samples obtained by means of AFM and TEM are reported. As seen, while TEM reveals the mean inorganic core diameter, the AFM images include also the coating thickness. Moreover, from the inset of Fig. 2 it can be noted that the cores of the particles are almost spherical.

The ZFC curves for samples A and B (a representative example for sample B is shown in Fig. 2), present a broad peak which suggests a distribution of MNPs sizes. The temperature of the maximum of this peak ($T_{max}$), is related to the average blocking temperature, which determines the transition from the superparamagnetic to the blocked state. The temperature $T_{max}$ of the ZFC curve for sample C could not be detected in the investigated temperature range. The hysteresis loops acquired at 2.5 K for the three samples are shown in Fig. 3. The parameters that characterize the magnetic behavior of the MNPs, i.e. the saturation magnetization $M_s$ and the coercivity field $H_c$, are listed in Table I.

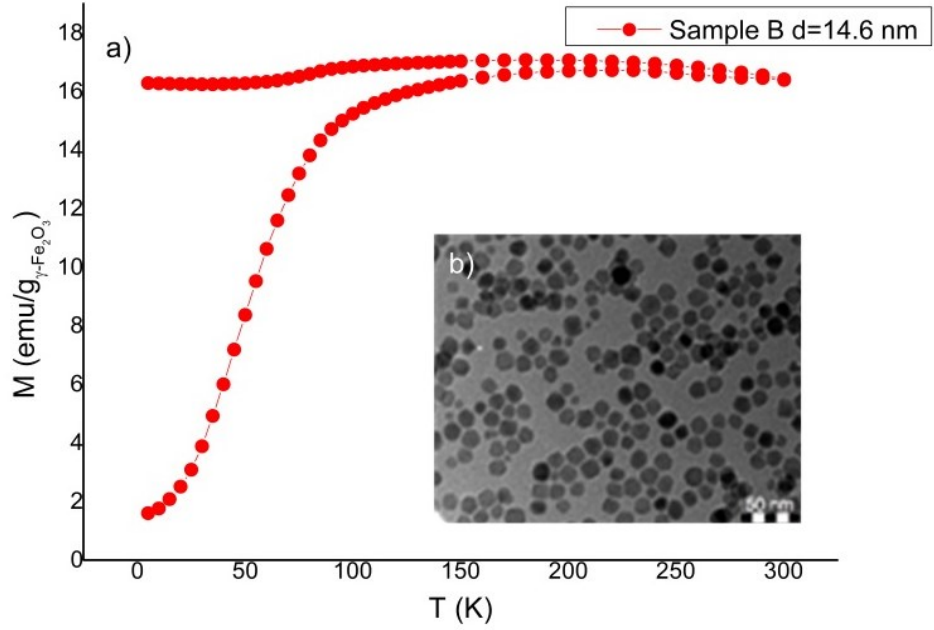

FIG. 2. (a) Temperature dependence of the ZFC/FC magnetization for sample B ($d$ = 14.6), collected at $\mu_0 H$ = 5 mT. In the inset (b), a TEM image of this sample is reported.

| Sample | $d_{TEM}$ (nm) | $d_{AFM}$ (nm) | $M_s$ at 2.5K (emu/g) | $H_c$ (Oe) | $M_s$ at 300K (emu/g) |
|---|---|---|---|---|---|
| A | 10.2±1.1 | 11.4±0.9 | 62.4±3.4 | 265±13 | 54.6±3.0 |
| B | 14.6±1.8 | 15.6±0.8 | 67.2±3.7 | 239±15 | 58.3±3.2 |
| C | 19.7±1.7 | 20.5±0.8 | 69.3±3.8 | 360±12 | 60.9±3.3 |

TABLE I. Diameter mean values and standard deviation of the MNPs samples as evaluated from statistical analysis with TEM (first column) and AFM (second column). The saturation magnetization $M_s$ at $T$ = 2.5K and $T$=300K and the coercive field $H_c$ at $T$ = 2.5 K are also reported.

Magnetic heating experiments were performed on stable aqueous solutions of MNPs at room temperature by a Magnetherm (nanoTherics$^{TM}$) set-up, using coils with 9 or 17 turns (depending on the operating field and frequency), and a diameter of 4.4 cm. This apparatus allows to vary by about one order of magnitude the AMF frequency $f$ (100 kHz - 1 MHz) and amplitude $\mu_0 H$ (3 kA/m - 17 kA/m). For each measurement, approximately 1 ml of MNP suspension was placed into a polystyrene thermally insulating sample holder in the middle of the inductive coil.

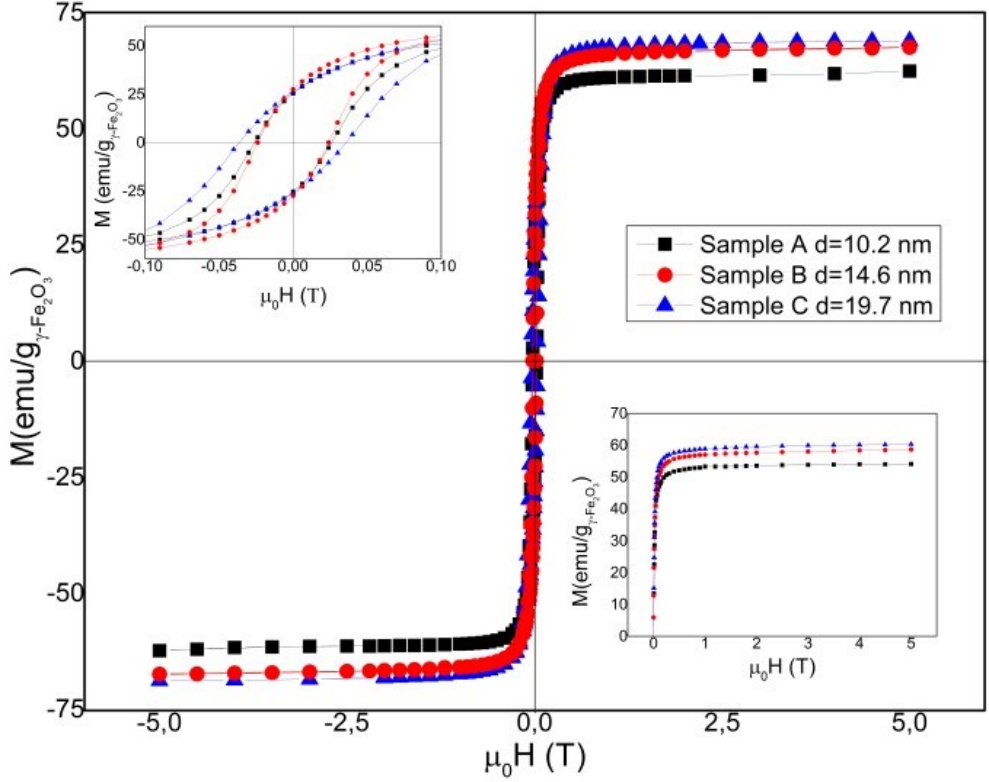

FIG. 3. Hysteresis loops measured at T = 2.5 K. In the top inset, a zoom highlights the residual magnetization and coercive field; in the bottom inset the first magnetization curves at 300 K are reported. *d* is the core diameter.

The temperature was measured using an optical fiber thermometer (Optocon$^{TM}$), dipped into the magnetic suspension, and the temperature vs. time curve, *T(t)*, was acquired during an interval of 5÷15 minutes after the AMF switching on. The *SLP* value was evaluated using the following equation:

$$SLP = \frac{m_{H_2O} c_{H_2O} + m_{\gamma-Fe_2O_3} c_{\gamma-Fe_2O_3}}{m_{\gamma-Fe_2O_3}} \frac{\Delta T}{\Delta t} \qquad (1)$$

where $c_{H2O}$ and $c_{\gamma-Fe2O3}$, $m_{H2O}$ and $m_{\gamma-Fe2O3}$ are, respectively, the specific heat and the mass of the main constituents of the solution (water and γ-Fe$_2$O$_3$ MNPs core, the coating contribution being neglected), and *t* is the time [31]. In the calculation, the specific heat was assumed $c_{H2O}$ = 4.2 JK$^{-1}$g$^{-1}$ for water, and $c_{\gamma-Fe2O3}$ = 0.61 JK$^{-1}$g$^{-1}$ for the maghemite core. To estimate the temperature increment rate *ΔT/Δt*, we mainly used the initial slope method, by fitting the first 15÷20 seconds of the *T(t)* curve. However, when the temperature increase was very low, we extracted the initial slope from the best fit of the whole *T(t)* curve to the Box-Lucas model *T(t) = α(1 − e$^{-βt}$)*, where *α* and *β* are parameters which represent the saturation temperature and the curvature of the heating curve, respectively. The *SLP* value (proportional to the *ΔT/Δt* ratio) used in Eq. (1) can be obtained by the product *α·β* [32,33].

The *SLP* values evaluated by means of Eq. (1) are reported in Fig. 4 for all the samples. The *SLP* of sample C is the highest at all measured frequencies and amplitudes, while samples A and B show slight differences, being the *SLP* values of B tendentially larger than the ones of A. In all systems, the amount of energy ab-sorbed from the AMF and then released as heat during one cycle of the applied field is proportional to the area $A$ of the hysteresis loop, so that the *SLP* is given by $SLP(W/g) = f \cdot A/\rho$ where $f$ is the frequency of the measurement and $\rho$ is the concentration of MNPs in the sample. The expression of $A$ can be analytically determined [14] for single domain nanoparticles for which the LRT can be applied (see Fig. 1).

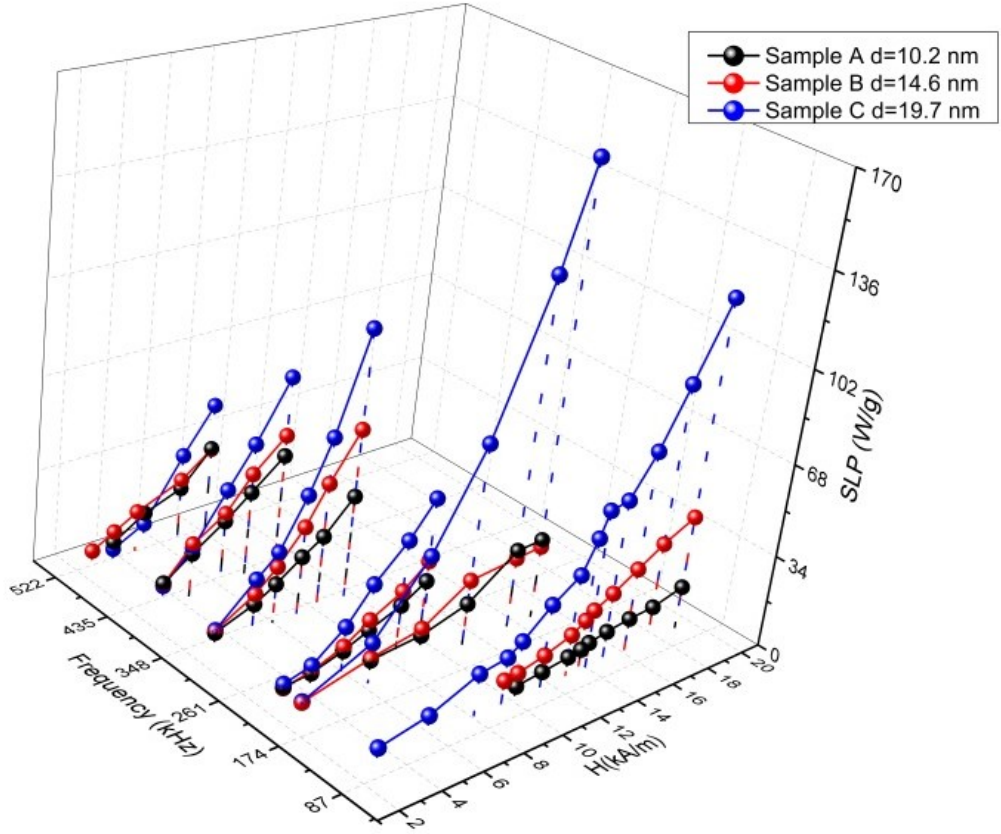

FIG. 4. *SLP* values for different frequencies and amplitudes of the alternating magnetic field. $d$ is the core diameter of the MNPs.

According to the LRT, indeed, the *SLP* can be written [34,35]:

$$SLP = \frac{\mu_0 \pi \chi''(f) f H_{max}^2}{\rho} \qquad (2)$$

where $H_{max}$ is the field amplitude and $\chi''$ is the out-of-phase component of the magnetic susceptibility:

$$\chi''(f) = \frac{\mu_0 M_s^2 V}{3 k_B T} \frac{2\pi f \tau_{eff}}{\left[1 + (2\pi f \tau_{eff})^2\right]} \qquad (3)$$

having divided by the factor 3 because of the MNPs random orientation [25] and indicating with $\tau_{eff}$ the effective relaxation time of MNPs, defined by $1/\tau_{eff} = 1/\tau_N + 1/\tau_B$. It should be noted that $1/\tau_{eff}$ is the correlation time corresponding to the fastest process between the Néel and Brown relaxation, if their characteristic time-scales are very different. According to Eq. (1)÷(3), when $2\pi f\tau_{eff} \ll 1$ (fast relaxation), the losses increase with the square of frequency and field, i.e. the *SLP* is proportional to $f^2\tau_{eff}H^2$, while for $2\pi f\tau_{eff} \gg 1$ (slow relaxation) the losses approach a frequency independent saturation value, i.e. the *SLP* is proportional to $H^2/\tau_{eff}$. When $2\pi f\tau_{eff} \approx 1$, i.e. as the transition from SP to FM regime is approached (Fig. 1), the LRT progressively fails by increasing the particles diameter and the SWM can be applied [36].

To verify experimentally which model is appropriate for our systems, we introduced in $\xi$ the values of $M_s$ estimated from the magnetic measurements and the volume $V$ obtained by TEM, and then we calculated for each sample the maximum value of the field $H$ that satisfies the condition $\xi < 1$ for the applicability of the LRT model. The resulting values of $H_{max}$, reported in Fig. 5a (coloured shadowed zones), highlight the values below which the condition of validity for the LRT is satisfied, showing that this model applies only for sample A. These observations have been confirmed by fitting the *SLP* data at 110 kHz to a power law with a free exponent, $SLP \propto H^x$. As expected, the best-fitting curves (Fig. 5(a), continuous lines) show that only the data of sample A can be described by a $H^2$ dependence at all frequencies. On the contrary, for B and C the differences between the experimental points and the $H^2$-behavior (dotted lines) are evident, particularly at high fields. From the fitting procedure, the exponent values for sample B and C are $x = 2.23 \pm 0.08$ and $x = 2.3 \pm 0.1$, respectively. Similar results were already reported [37]. According to the theoretical framework introduced above, intermediate values between 2 and 3 of the exponent means that the MNPs fall in the transition region SP-FM.

The same function $SLP \propto H^x$ was used to fit the data at 237 kHz and 340 kHz (solid lines) in the magnetic field region $\mu_0 H \leq 10$ kA/m (upper field limit for our experimental apparatus at high frequencies) and the $x$ values are reported in Table II. While $x$ remains almost constant and equal to 2 (within 5%) for A, it increases for B and, mainly, C with increasing frequency. This result can be explained considering that the transition SP-FM occurs when $2\pi f\tau_{eff} \approx 1$ and, being $\tau_{eff}$ constant for each sample, when the frequency is increased the system tends to the FM (or blocked) regime. According to this interpretation, sample A is clearly far from the transition, while B and C progressively block in the FM regime.

| Sample | $x$ at 110 kHz | $x$ at 237 kHz | $x$ at 340 kHz |
|---|---|---|---|
| A | 2.0±0.1 | 2.03±0.01 | 2.09±0.09 |
| B | 2.23±0.08 | 2.30±0.02 | 2.34±0.03 |
| C | 2.3±0.1 | 2.47±0.09 | 2.64±0.06 |

TABLE II. Free exponent $x$ of the function $SLP \propto H^x$ for the samples at three different frequencies.

As a final step, we evaluated the effective relaxation time for the magnetization reversal from the *SLP* data to check the validity of the LRT phenomenological model. On this respect, it should be also noted that we could get the effective relaxation time $\tau_{eff}$ at different frequencies only for sample A, for which the *SLP* is analytically expressed [Eq. (2) and (3)]. Thus, rewriting Eq. (2) in the form $SLP = \delta \cdot H^2$, from the fits of the SLP($H$) curves at a given frequency we obtained the values of the coefficient $\delta$ and, reversing Eq. (3), the relaxation time $\tau_{eff}$. Using this method for $f$ = 110 KHz, we found $\tau_{eff}$ = 19.5 ± 2.3 ns. This value can be compared with the one obtained from magnetic data. By using for the anisotropy constant $K$ = 1.8·10$^4$ J/m$^3$, estimated from the hysteresis measurements at 2.5 K (from the formula $\mu_0 H_c$=0.96($K/M_s$)[1-($T/T_B$)$^{0.77}$]), and for the attempt time $\tau_0$ = 10$^{-9}$ s, we obtained $\tau_N$= 17.6 ± 2.5 ns. Comparing this value with the one obtained from the fit of the *SLP* vs $H$ curve at 110 kHz, we can observe a good agreement within the experimental error. As a further consequence, we can assert that for sample A the main mechanism involved in the magnetic relaxation of the nanoparticles is the Néel process, while the Brown rotation has a negligible role. This is confirmed by the rough estimation of $\tau_B$ obtained by using the diameter deduced from AFM, from which $\tau_N$ << $\tau_B \approx$ 890 ± 20 ns, and thus $\tau_{eff} \approx \tau_N$.

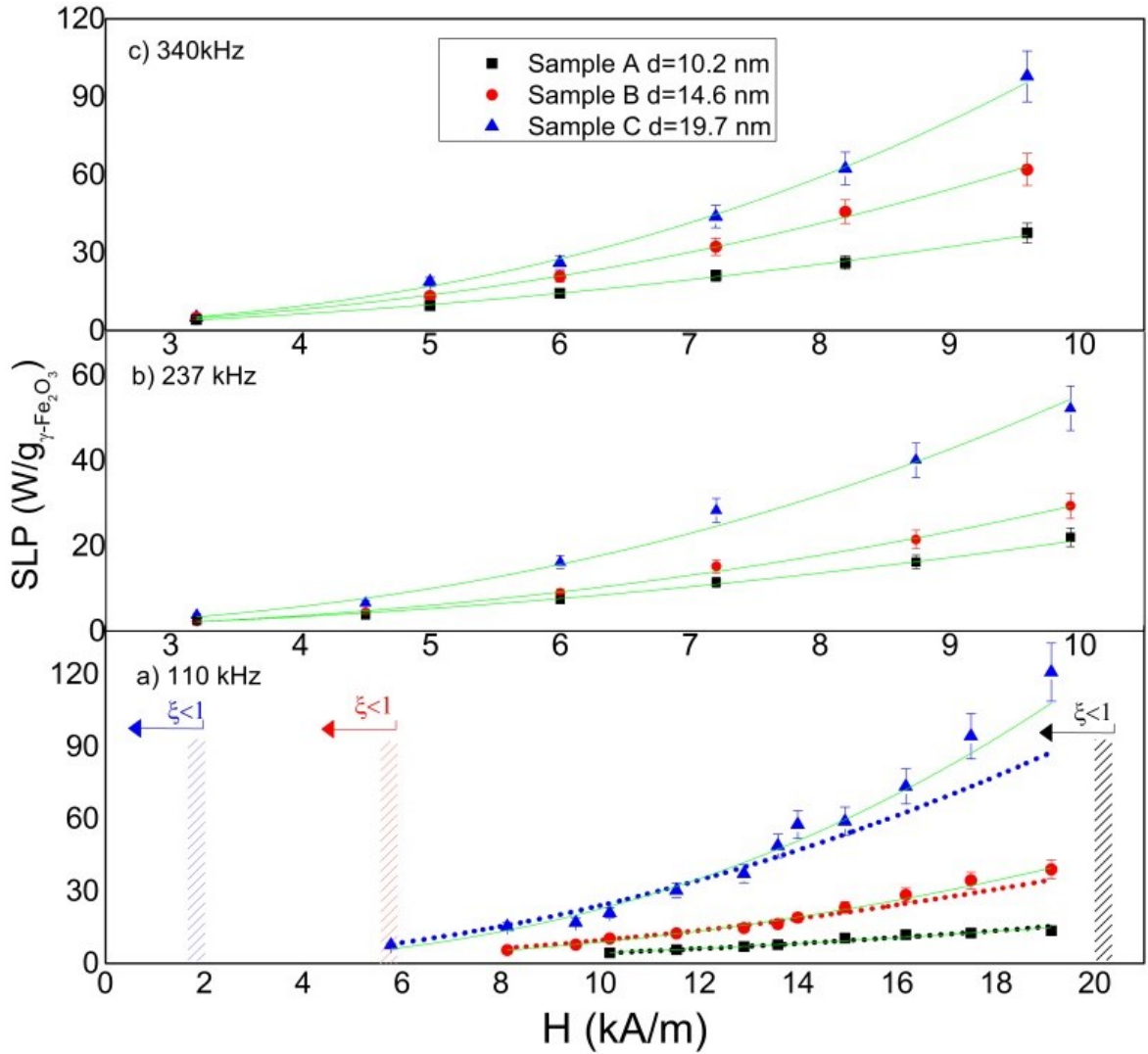

FIG. 5. *SLP* vs. AMF amplitude at 110 kHz (a), 237 kHz (b) and 340 kHz (c). The dotted lines represent the prediction of the LRT model, $SLP \propto H^2$, and the continuous lines the fitting function $SLP = c \cdot H^x$ (2 < $x$ < 3). The coloured shadowed regions individuate the upper limits (see also arrows) of validity of the LRT model for the corresponding samples.

## IV. CONCLUSIONS

In conclusion, we systematically investigated the heating properties of $\gamma$-$Fe_2O_3$-based MNPs with three different diameters, when excited by an alternating magnetic field. The experimental results were collected over a wide range of field frequencies and amplitudes, at an extent not found in the literature. While the *SLP* of MNPs grows up as the diameter increases, its *H*-dependence changes as a function of the MNPs core diameter and the applied field frequency. The LRT model, where *SLP* is proportional to $H^2$, explains the data for small MNPs (sample A, $d = 10.2$ nm), but fails when the MNP size ($d = 14.6$ and $19.7$ nm for samples B and C, respectively) becomes larger than the critical diameter corresponding to the onset of the transition from the SP to the FM regime, as expected. In particular, for samples B and C the *SLP* is no longer proportional to $H^2$ but follows the power law $SLP = c \cdot H^x$ with $2 < x < 3$, where $x$ increases with increasing frequency. These $x$ values are intermediate between the one predicted by the LRT ($x = 2$) and the Rayleigh ($x = 3$) models. Our experimental results confirm the theoretical predictions [25,26] concerning the evolution of the $x$ exponent with the MNPs dimensions. Finally, the validity of the LRT model for sample A allowed us to directly evaluate the effective relaxation time from the *SLP* best-fitting curve and, by comparison with the estimation of $\tau_{eff}$ from the hysteresis curve, to establish that the dominant relaxation is given by the Néel process.

**ACKNOWLEDGMENT**

The COST-RADIOMAG from EU and Fondazione Banca del Monte di Lombardia are acknowledged for partly funding this research.

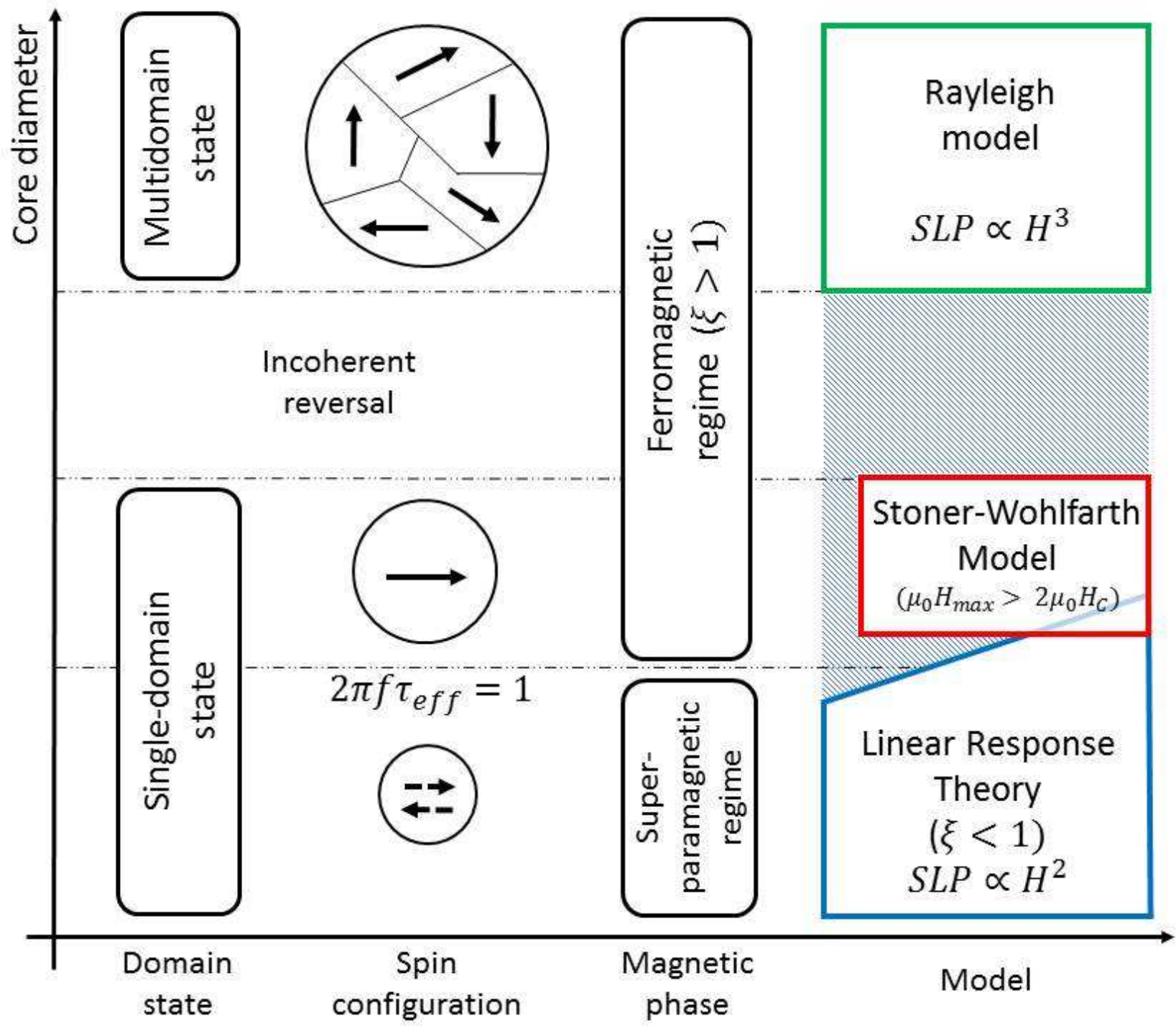

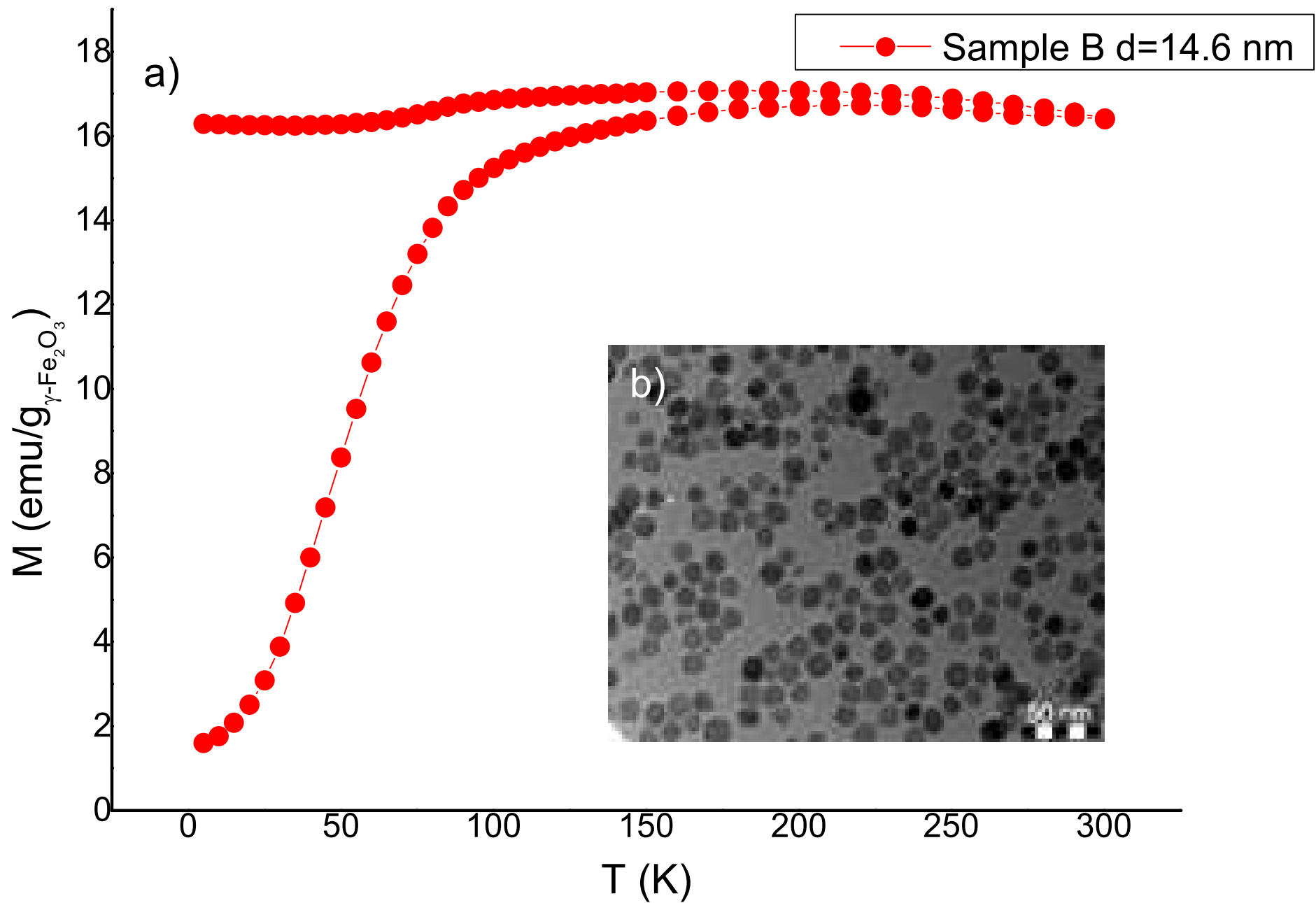

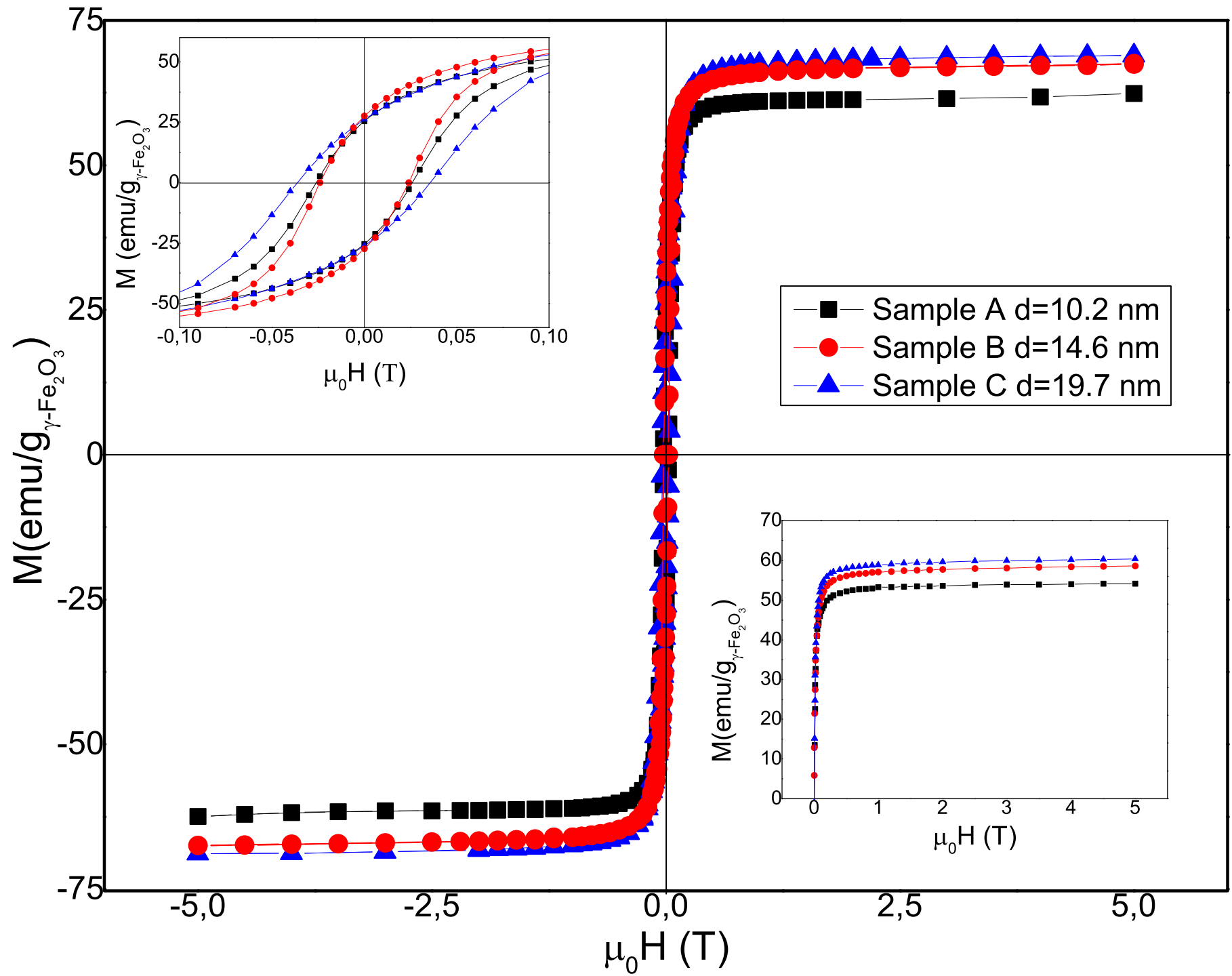

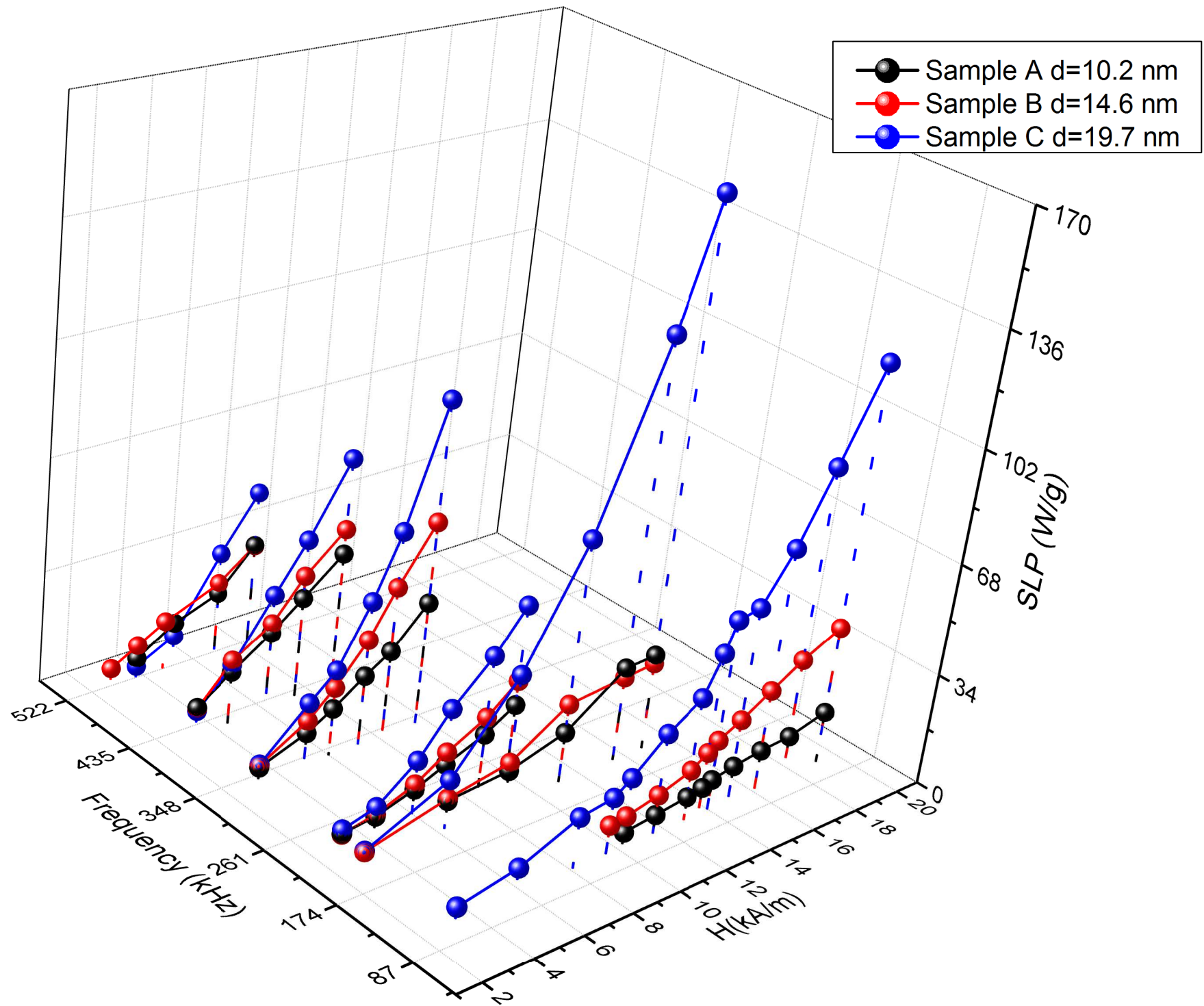

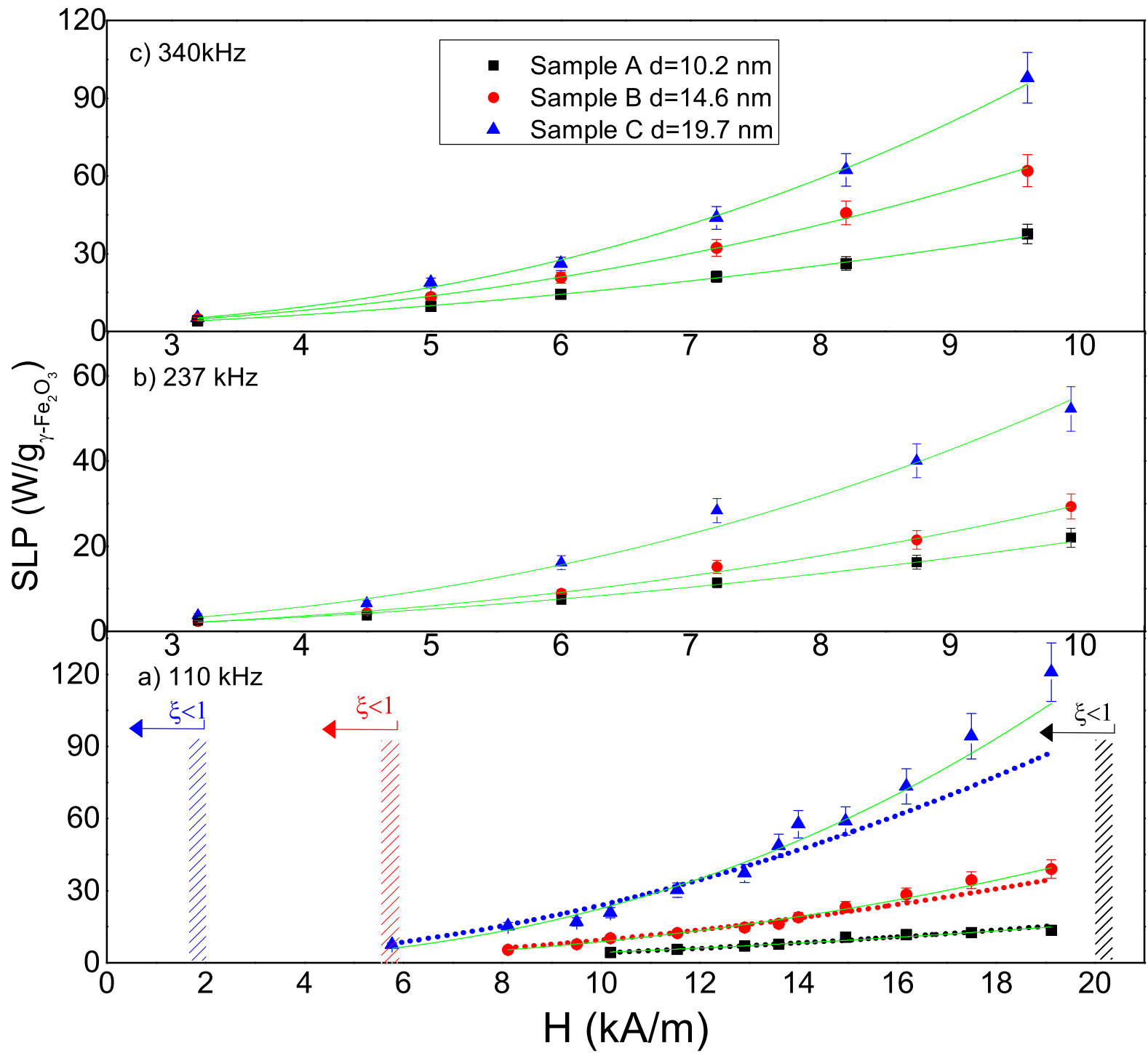